\documentclass[a4paper,11pt]{article}
\usepackage{amsfonts}
\newlength{\dinwidth}
\newlength{\dinmargin}
\newtheorem{theorem}{Theorem}
\newtheorem{corollary}{Corollary}
\newtheorem{lemma}{Lemma}
\newtheorem{proposition}{Proposition}
\def\be{\begin{equation}}
\def\ee{\end{equation}}
\def\ben{\begin{displaymath}}
\def\een{\end{displaymath}}
\def\baa{\begin{eqnarray}}
\def\eaa{\end{eqnarray}}

\def\ba{\begin{array}}
\def\ea{\end{array}}
\def\Box{\diamond}
\def\C{\mathbb{C}}
\def\Z{\mathbb{Z}}
\def\R{\mathbb{R}}
\makeatletter
\@addtoreset{equation}{section}
\makeatother

\def\a{\alpha}

\def\b{\beta}

\def\l{\lambda}

\def\Th{\Theta}
\def\o{\omega}

\def\xibar{{\overline{\xi}}}
\def\Thh{\Theta_{\pb\qb}}
\def\Thho{\Theta_{\pb\qb}(\int_\xi^{\infty^+})}
\def\Thhd{\Theta_{\pb\qb}(\int_\xi^{\infty^-})}
\def\Thhxi{\Theta_{\pb\qb}(\int_\xi^{\xibar})}
\def\Thhdb{\Theta_{\pb\qb}(\int_\xibar^{\infty^-})}

\def\Thxi{\Theta(\int_\xi^{\xibar})}

\def\co{c_1(\infty^-,\xi,\infty^+)}
\def\coxi{c_1(\xi,\xibar,\infty^-)}
\def\cdxi{c_2(\xi,\xibar,\infty^-)}
\def\cd{c_2(\infty^-,\xi,\infty^+)}

\def\dd{d_2(\xibar,\xi)}

\def\realpart{Q}
\def\Ec{{\cal E}}

\def\B{{\bf B}}
\def\CP1{\C{\mathbb{P}} 1}
\def\z{{\bf z}}

\def\la{\label}

\def\f{\frac}
\def\L{{\cal L}}
\def\p{\partial}
\def\pb{{\bf p}}
\def\qb{{\bf q}}

\def\mb{{\bf m}}

\def\zb{{\bf z}}

\def\0{S}
\def\log{\ln}

\def\B{{\bf B}}
\def\la{\label}

\def\f{\frac}
\def\L{{\cal L}}
\def\p{\partial}

\def\0{S}
\def\1{T}

\def\log{\ln}

\def\bar{\overline}
\def\det{{\rm det}}

\def\V{{\bf V}}
\def\mb{{\bf m}}
\begin{document}

\title{Ernst equation, Fay identities and variational formulas \\ on hyperelliptic curves}
%\shorttitle{Riemann-Hilbert problems and branched coverings}
\maketitle
\vskip0.3cm
\begin{center}
{\Large C.Klein$^1$, D.Korotkin$^2$ and V.Shramchenko$^2$}\\
\vskip0.2cm
$^1$ Max-Planck-Institut f\"ur Physik, F\"ohringer 
Ring 6, 80805 M\"unchen, Germany\\
$^2$Department of Mathematics and Statistics, Concordia University\\
7141 Sherbrook West, Montreal H4B 1R6 Quebec, Canada
%\{korotkin@mathstat.concordia.ca}
\end{center}
\begin{abstract}
We present a unified approach to theta-functional solutions of the 
stationary axisymmetric Einstein equations in vacuum. Using Fay's 
trisecant identity and variational formulas on hyperelliptic Riemann 
surfaces, we establish formulas for the metric functions, the Ernst 
potential and their derivatives. 
\end{abstract} 

\section{Introduction}

The theory of theta-functional solutions of integrable equations starts 
in 1974 with  the works of
Novikov, Dubrovin, Matveev, Its and Krichever (see 
\cite{DubNov75,ItsMat75,Kric78} 
and references therein) on periodic and 
quasi-periodic solutions to the  Korteweg de Vries (KdV) equation.
These finite-gap solutions can be expressed via the Its-Matveev formula \cite{ItsMat75} in terms of 
a second derivative of multidimensional theta-functions of a hyperelliptic 
algebraic curve. The main technical tool exploited in \cite{DubNov75,ItsMat75} was the spectral theory of Sturm-Liouville 
operators with periodic potential. Later Krichever \cite{Kric78}
generalized the Its-Matveev formula to an integrable  
generalization  of the KdV equation to 2+1 dimensions - the  Kadomtzev-Petviashvili (KP) equation. 
Analogous formulas in terms of theta-functions  were derived in the framework of the 
inverse scattering method for other integrable equations as, for example,
Sine-Gordon, Non-linear Schr\"odinger, and Landau-Lifschitz.

On the other hand  \cite{Shiota,Mumford}, the finite-gap solutions of
 integrable systems of KP type may
 be derived directly from  Fay's trisecant  identity \cite{Fay73}.   
This identity holds for any set of four points 
$(a,b,c,d)$  on a compact Riemann surface $\L$ of genus $g$ and any vector $\z\in \C^g$:
\ben
 E(c,a)E(d,b)\Th(\z+\int_b^c)\Th(\z+\int_a^d)+E(c,b)E(a,d)\Th(\z+
\int_a^c)\Th(\z+\int_b^d)
\een
\ben
=E(c,d)E(a,b)\Th(\z)\Th(\z+\int_b^c+\int_a^d)\;,
\een
where $\Th$ is the theta-function built from the matrix of $b$-periods of the surface $\L$; $\int_b^a$ is a short notation for the difference of Abel 
maps on   $\L$ between  points $a$ and $b$; $E$ is the prime-form on
 $\L$ (see sect.2.1).

Further development of the method of finite-gap integration of integrable systems in \cite{Koro88} allowed to solve in  terms 
of theta-functions the  Ernst equation
\be
(\Ec+\bar{\Ec})(\Ec_{\zeta\zeta}+\f{1}{\rho}\Ec_\rho+\Ec_{\rho\rho})=2(\Ec_\zeta^2+\Ec_\rho^2)
\la{Ernstint}\ee
for a complex-valued Ernst potential $\Ec$ depending on two coordinates $(\zeta,\rho)$. The Ernst equation is equivalent to the
stationary axially symmetric  vacuum Einstein equation; it was embedded in the framework of the inverse scattering method by 
Belinski-Zakharov \cite{belzak} and Maison \cite{maison} in 1978. In particular, certain multisoliton solutions of the Ernst equation (which form a degenerate subclass
of algebro-geometric solutions) give rise to Schwarzschild and Kerr black holes. 
A class of non-degenerate theta-functional solutions of (\ref{Ernstint})  was recently used in 
\cite{NeuMei95,Klein00} to describe the gravitational field of rotating dust discs.

The  theta-functional solutions of (\ref{Ernstint}) can be written in the 
form \cite{KorMat00}
\be
\Ec=\frac{\Thho}{\Thhd},
\la{solint}\ee
where the theta-function corresponds to the hyperelliptic spectral curve
$\mu^2=(\l-\xi)(\l-\xibar)\prod^g_{k=1}(\l-E_k)(\l-F_{k})$, where  
$\xi=\zeta+i\rho$, and where for each $k$ we require that either 
$E_k=\bar{F_k}$ or $E_k,F_k\in \R$.
The constant (with respect to the physical 
coordinates) vectors $\pb$ and $\qb$ must satisfy the
reality condition $\B\pb+\qb\in \R^g$. The notations  $\infty^+$ and $\infty^-$ are used for the infinite points on different sheets of the curve $\L$, namely, $\mu/\l^{g+1}\to \pm 1$ as $\l\to\infty^\pm$, respectively.   

We notice the essential difference between the solution (\ref{solint})
of the Ernst equation, and, say, the 
finite-gap solutions  of the  KP equation.
The spectral curve of the Ernst equation is ``dynamical'' i.e. it depends on the space-time coordinates $(\xi,\xibar)$, whereas the 
spectral curve of the KP equation is static, i.e. it is built from the integrals of motion of the system. 
The dynamical character of the spectral curve of the Ernst equation 
implies, in particular, the asymptotical flatness of the theta-functional solutions, in contrast to 
the quasi-periodic nature  of previously known theta-functional
solutions of equations of KdV and KP type. The solutions 
(\ref{Ernstint}) thus are defined on a family $\L(\xi,\bar{\xi})$ of 
Riemann surfaces and are in general not periodic or quasi-periodic.

The original derivation of the solutions (\ref{solint}) was performed 
with the use of the zero curvature representation of the Ernst equation,
and the  solution of an appropriate Riemann-Hilbert problem. The explicit formulas for the  coefficients 
of the space-time metric corresponding to solutions (\ref{solint}) required the calculation of the tau-function corresponding to this
Riemann-Hilbert problem.

It is the purpose of the present paper to prove the formulas for the 
Ernst potential  (\ref{solint}) and the metric functions, 
by using only the trisecant Fay identity (\ref{Fay})
(together with its appropriate degenerations) and  Rauch's variational 
formulas \cite{Rauch} which describe the  dependence of
the holomorphic differentials on the moduli of the Riemann surface. 
Therefore we confirm once more the universality of Fay's identities
in the theory of integrable systems, and show their applicability to non-autonomous equations of Ernst type.

In section 2 we collect some useful facts from the theory of Riemann surfaces. In section 3 we prove, 
using  Fay's identity  and   Rauch's formulas, that the function   (\ref{solint}) satisfies
the Ernst equation. Finally, in section 4 we use the same techniques 
for formulas for the metric coefficients corresponding to this solution 
of the Ernst equation. Section 5 contains a summary and an outlook.

\section{Fay's identities and Rauch's variational formulas}

\subsection{Fay's identities and their degenerations}

Consider a compact Riemann surface $\L$ of genus $g$.
On this surface we introduce a canonical basis of cycles $(a_\a, 
b_\a)$, $\a=1,\dots,g$, the basis of holomorphic differentials normalized 
by the condition $\int_{a_\a}\o_\b=\delta_{\a\b}$,
and the matrix $\B_{\a\b}=\int_{b_\a}\o_\b$ of their $b$-periods.
The theta-function with characteristics corresponding  to the curve $\L$ 
is given by
\be
\Thh(\z|\B)=\sum_{\mb\in \Z^g}\exp\{\pi i\langle\B(\pb+\mb),(\pb+\mb)\rangle +
2\pi i \langle\pb+\mb,\qb+\z\rangle\};
\la{theta}\ee
here $\z\in\C^g$ is the argument and $\pb,\qb\in\C^g$ are the vectors of 
characteristics; $\langle.,.\rangle$ denotes the
scalar product.
The theta-function with characteristics is, up to an exponential factor, equal to the theta-function with 
zero characteristics (denoted by $\Theta$)
and shifted argument:
\be
\Thh(\z|\B)= \Th(\z+\B\pb+\qb) \exp\left\{\pi i \langle\B\pb,\pb\rangle+ 2\pi i\langle\pb,\z+\qb\rangle\right\}
\la{thchar}\ee
The theta-function satisfies the  heat equation: 
\be
4\pi i\partial_{\B_{\a\b}}\{\Thh(\z,\B)\}=\partial_{z_\a}\partial_{z_\b}
\Thh(\z,\B).
\la{heat}\ee

The main tool we are going to exploit in this paper is Fay's trisecant identity for theta functions and prime forms. The prime form is the $(-\frac{1}{2},-\frac{1}{2})$-differential on $\L\times\L$
given by 
$$E(a,b)=\frac{\Th_{\star}(\int_b^a)}{h_\Delta(a)h_\Delta(b)},$$ 
where 
$h_\Delta^2(a)=\sum_{\a=1}^g\frac{\partial\Th_{\star}}{\partial z_\a}(0)\o_\a(\tau_a)$, 
and where $\star\equiv [\pb^\star\qb^\star]$ is an odd non-singular  half-integer characteristic (note that 
the prime form is independent of the choice of the 
characteristic  $\star$). As before,
$\int_a^b$ denotes the line integral from $a$ to $b$ of the vector 
$\o(\tau)=(\o_1(\tau),\dots,\o_g(\tau))^T$.

Fay's trisecant identity holds for any four  points $ a,b,c,d\in\L$ and any two characteristic vectors $\pb,\qb\in\C^g$:
\ben E(c,a)E(d,b)\Thh(\z+\int_b^c)\Thh(\z+\int_a^d)+E(c,b)E(a,d)\Thh(\z+
\int_a^c)\Thh(\z+\int_b^d)
\een
\be
=E(c,d)E(a,b)\Thh(\z)\Thh(\z+\int_b^c+\int_a^d).
\la{Fay}\ee 
where all integration contours are chosen not to intersect the canonical basic cycles; this requirement completely
fixes all terms of the identity (\ref{Fay}).
 
In the sequel we will use the degenerate versions  of  Fay's identity. 
Let us denote by $D_a$ the operator for the directional derivative along 
the basis of holomorphic differentials, 
 acting on theta-functions: 
\ben D_a\Thh(\z)=\langle \nabla \Thh(\z), \frac{\o(a)}{d\tau_a}\rangle\equiv\sum_\a\partial_{z_\a}\{\Thh(\z)\}\frac{\o_\a(a)}{d\tau_a}. 
\een
Since the theta-function (\ref{theta}) depends only on the sum of 
vectors $\z$ and $\qb$, the action of the operator $D_a$ on a
theta-function with characteristics can be written alternatively as follows:
\be
D_a\Thh(\z)=\sum_\a\partial_{q_\a}\{\Thh(\z)\}\frac{\o_\a(a)}{d\tau_a}
\la{defD}\ee
This form of $D_a$ can be easily  extended to any object depending on 
a vector $\qb$.

Differentiating (\ref{Fay}) with respect to the argument $d$ and taking the limit $d\to b$ one obtains 
\begin{corollary}
The following degenerated version of Fay's identity holds:
\be
D_b\log\frac{\Thh(\z+\int_a^c)}{\Thh(\z)}=c_1(a,b,c)
+c_2(a,b,c)\frac{\Thh(\z+\int_a^b)\Thh(\z+\int_b^c)}{\Thh(\z)\Thh(\z+\int_a^c)}\;,
\la{Fay1}\ee
where the functions of three variables $c_1$ and $c_2$ are given by:
\be
c_1(a,b,c)=\frac{\o_{a,c}(b)}{d\tau_b}, 
\la{c1}
\ee
where $\o_{a,c}(b)$
is the differential of the third kind with poles in $a,c$, and 
\be
c_2(a,b,c)=\frac{E(a,c)}
{E(a,b)E(b,c)d\tau_b}.
\la{c2}\ee
\end{corollary}
The derivative of (\ref{Fay1}) with respect to argument $c$ gives in the 
limit $c\to a$ 
\begin{corollary}
The following twice degenerated version of Fay's identity holds:
\be
D_aD_b\log\Thh(\z)=d_1(a,b)+d_2(a,b)\frac{\Thh(\z+\int_b^a)\Thh(\z+\int_a^b)}{\Thh^2(\z)},
\la{Fay2}\ee
where the functions of the two variables $d_1$ and $d_2$ are given by:
\be
d_1(a,b)=-\frac{W(a,b)}{d\tau_a d\tau_b}\;,
\la{d1}\ee
\be
d_2(a,b)=\frac{1}{E^2(a,b)d\tau_ad\tau_b}\;;
\la{d2}\ee
 $W(a,b)=d_{a}d_b\log E(a,b)$ is the Bergmann kernel.
\end{corollary}

\subsection{Root functions and Rauch's variational formulas on hyperelliptic curves} 
Let us now choose  $\L$ to be the  hyperelliptic 
algebraic curve with $2g+2$ branch points  defined by the equation
\be
\mu^2=\prod_{m=1}^{2g+2}(\l-\l_m)\;.
\la{hyper}
\ee

The following identity for  ``root functions'' holds \cite{Fay73} for any point $a\in\L$:
\be
\frac{E(a,{\l_m})\sqrt{d\tau_{\l_m}}}{E(a,{\l_n})\sqrt{d\tau_{\l_n}}}
=C\sqrt{\frac{\l(a)-{\l_m}}{\l(a)-{\l_n}}},
\la{root}\ee
where $\l(a)$   denotes the  projection of point $a$ onto the 
Riemann sphere; $C$  is a constant with respect to $\l(a)$.

Rauch's variational formulas \cite{Rauch} describe the dependence of the basic normalized 
holomorphic differentials $\o_\a$ and the matrix of $b$-periods $\B_{\a\b}$
 on the moduli of the
Riemann surface. The moduli space of hyperelliptic curves can be 
parameterized by the positions of the branch points, and Rauch's formulas 
read:
\be 
\frac{d\o_\a}{d\l_m}(a)=\frac{1}{2}\frac{W(a,\l_m)}{d\tau_{\l_m}}\frac{\o_\a(\l_m)}{d\tau_{\l_m}},
\la{Rauch1}
\ee
\be
\frac{d\B_{\a\b}}{d\l_m}=\pi i\frac{\o_\a(\l_m)}{d\tau_{\l_m}}\frac{\o_\b(\l_m)}{d\tau_{\l_m}},
\la{Rauch2}
\ee

The formulas (\ref{Rauch1}), (\ref{Rauch2}), together with the heat 
equation for theta-functions (\ref{heat}), imply the 
following dependence of hyperelliptic  theta-functions on the branch points:
\begin{lemma}
The derivative of the hyperelliptic  theta-function  $\Thh(\z)$ with a 
$\{\l_m\}$-dependent argument $\z$  with respect to a branch point   $\l_m$ is given by
\be
\partial_{\l_m}\Thh(\z)=\frac{1}{4}D_{\l_m}D_{\l_m}  \Thh(\z)+\sum_\a\partial_{z_\a}\{\Thh(\z)\}\frac{d\z_\a}{d\l_m}
\la{heat1}\ee
\end{lemma}

\section{Ernst equation and Fay identities}

Consider a real  hyperelliptic Riemann surface $\L$ of genus $g$ given 
by 
\be
\mu^2=(\l-\xi)(\l-\xibar)\prod^g_{m=1}(\l-E_m)(\l-F_m),
\la{L}\ee
where  $\xi=\zeta-i\rho$; $\zeta,\rho\in\R$; for each $m$ we require that either $E_m=\bar{F_m}$ or $E_m,F_m\in \R$.
Let us introduce the canonical basis of cycles on $\L$ according to Fig.1: the $a$-cycles are chosen to encircle the branch cuts
$[E_m,\;F_m]$; $b$-cycles all start at the branch cut $[\xi,\xibar]$.
In the sequel we  shall denote the  point which belongs to the upper sheet 
of $\L$ and has the projection $\l$ on $\CP1$ by $\l^+$;
the point which has the same projection on $\CP1$ but belongs to the lower sheet will be denoted by $\l^-$.

It is convenient to rewrite the Ernst equation (\ref{Ernstint}) in 
terms of the complex coordinates $(\xi,\xibar)$ as follows:
\be
(\Ec+\bar{\Ec})\left(\Ec_{\xi\xibar}-\frac{1}{2(\xibar-\xi)}(\Ec_\xibar-\Ec_\xi)\right)=2\Ec_\xi\Ec_\xibar\;.
\la{ErE}\ee

This section will be devoted to the proof of the following theorem using Fay's identities and Rauch's formulas:
\begin{theorem}
Let the branch points $E_m$, $F_m$ of the curve $\L$ (\ref{L}) be $(\xi,\xibar)$-independent.
Then the function 
\be
\Ec=\frac{\Thho}{\Thhd}\;,
\la{ES}\ee
where the theta-function corresponds to the matrix of $b$-periods of 
the curve 
$\L$, and where 
 an arbitrary $(\xi,\xibar)$-independent  non-singular  characteristic $[\pb,\qb]$ obeys the  reality conditions 
\ben
\B\pb+\qb\in \R^g\;,
\een
satisfies the Ernst equation (\ref{ErE}) in  the region of the $\xi$-plane, where the 
vector $\B\pb+\qb$ does not belong to the theta-divisor on the Jacobi
manifold of $\L$ (i.e. $\Thh(0)\neq 0$), and, in addition, 
\ben
\Thhd\neq 0\;.
\een
In accordance with the previous notation, $\int_\xi^a$ denotes the line integral
 of the vector $\o=(\o_1,\dots,\o_g)^T$  from $\xi$ to $a$. The integration paths in the numerator and denominator are supposed to have
the same projection onto $\CP1$; therefore, $\int_\xi^{\infty^+}=-\int_\xi^{\infty^-}$.
\end{theorem}
The proof will consist of a series of auxiliary statements: we shall  compute 
the derivatives of the  Ernst potential with 
respect to $(\xi,\xibar)$ and the action of the cylindrical Laplace operator
\be
\Delta\equiv\p^2_{\rho\rho}+\f{1}{\rho}\p_\rho+\p_{\zeta\zeta}^2\equiv
4\left(\p^2_{\xi\xibar}-\frac{1}{2(\xibar-\xi)}
\left(\p_\xibar-\p_\xi \right)\right)
\la{delta0}\ee
on the Ernst potential. 
We note that the real part of the Ernst potential can be written in a 
compact form: 
\begin{proposition}
The real part of the Ernst potential is given by the following expression:
\be
\Ec+{\bar{\Ec}}=2\realpart\frac{\Thh(0)\Thh(\int_{\xibar}^\xi)}{\Thhd\Thhdb}
\la{E+Ebar}\ee
where the function
\be
\realpart(\xi,\xibar)=\frac{1}{2}\frac{E(\xi,\xibar)E(\infty^-,\infty^+)}{E(\xi,\infty^-)E(\xibar,\infty^+)}
\la{C0}\ee
does not depend on $\pb,\qb$.
Taking into account that $\Ec\equiv 1$ if $\pb=\qb=0$, we get an 
alternative form of the function $\realpart$ in terms of theta-functions with
zero characteristics ($\Th \equiv \Theta_{\mathbf{0}\mathbf{0}}$):
\be
\realpart=\f{\Th(\int_\xi^{\infty^-})\Th(\int_\xibar^{\infty^-})}{\Th(0)\Th(\int_\xi^{\xibar})}\;.
\la{C00}\ee
\end{proposition}
{\it Proof.}
The proof is an immediate corollary of Fay's identity (\ref{Fay}) 
applied to the points $(\infty^+,\infty^-,\xi,\xibar)$ if we note the 
following 
\begin{lemma}\la{stralem}
The following relation holds:
\be
\frac{E(\infty^+,\xibar)E(\infty^-,\xi)}{E(\infty^-,\xibar)E(\infty^+,\xi)}=-1\;.
\la{strange}\ee
\end{lemma}
{\it Proof.}
To prove ({\ref{strange}) we use formula (\cite{Fay73}, p.21) which is 
valid for arbitrary four points $a,b,c,d$ on $\L$:
\be
\log\f{E(b,d)E(a,c)}{E(a,d)E(b,c)}=\int_c^d \omega_{b,a}\;,
\ee
where $\omega_{b,a}$ is normalized (all $a$-periods vanish) differential of the third kind on $\L$ with poles at $a$ and $b$ and residues $-1$ and $+1$,
respectively.
Assuming $a=\xibar$, $b=\xi$, $c=\infty^-$, $d=\infty^+$, we get the integral $\int_{\infty^-}^{\infty^+} \omega_{\xi,\xibar}$
along the path encircling the branch point $\xi$. On the hyperelliptic
curve (\ref{L})
with our choice of canonical cycles (Fig.1)
the abelian integral $\int \omega_{\xi,\xibar}$ can be computed
explicitly to give $\f{1}{2}\log\f{\l-\xi}{\l-\xibar}+C$, where $C$ is
an arbitrary constant (indeed, this expression has the required
structure of singularities at $\xi$ and $\xibar$, and does not suffer any modification with respect
to tracing along $a$-cycles shown in Fig.1; we remind that the local
parameters around $\xi$ and $\xibar$ are $\sqrt{\l-\xi}$ and
$\sqrt{\l-\xibar}$, respectively). Therefore, 
\ben
\int_{\infty^-}^{\infty^+} \omega_{\xi,\xibar}=\f{1}{2}\log\f{\l-\xi}{\l-\xibar}\Big|_{\infty^+}^{\infty^-}=\f{1}{2}2\pi i=\pi i\;,
\een
which gives (\ref{strange}).
$\Box$
$\Box$

\subsection{First derivatives of the Ernst potential}

We will first give convenient relations for the first derivatives 
of the Ernst potential which where obtained in \cite{Klein96} with the 
use of the zero-curvature representation of the Ernst equation.
\begin{proposition}
The first derivatives of the Ernst potential (\ref{ES}) are given by the following expressions:
\be
\Ec_\xi=\frac{c_2(\infty^-,\xi,\infty^+)}{2}
\frac{\Thh(0)}{\Thh^2(\int_\xi^{\infty^-})}D_\xi\Thh(0).
\la{Epoxi}\ee
\be
\Ec_\xibar=\frac{c_2(\infty^-,\xibar,\infty^+)}{2}
\frac{\Thh(\int_\xibar^\xi)}{\Thh^2(\int_\xi^{\infty^-})}D_\xibar
\Thh(\int_\xi^\xibar),
\la{Epoxibar}\ee
where $c_{2}$ is the constant (\ref{c2}) from the degenerated  Fay identity (\ref{Fay1}).
\end{proposition}
{\it Proof.}
Let us first note the following corollary of Rauch's variational  formulas:
\be
\frac{d}{d\xi}\int_\xi^{\infty^+}\o_\a(\tau)\equiv -\frac{d}{d\xi}\int_\xi^{\infty^-}\o_\a(\tau)=-\frac{1}{4}c_1(\infty^-,\xi,\infty^+)\frac{\o_\a(\xi)}{d\tau_\xi}.
\la{c10}\ee
where $c_1$ is as defined in (\ref{c1}).
To prove (\ref{c10}) we notice that, according to (\ref{Rauch1}), the derivative 
of a holomorphic differential with respect to a branch point
is proportional to the normalized differential of the second kind (the 
Bergmann kernel); consequently the integration of 
this differential gives a differential of the third kind, according to (\ref{c1}), (\ref{c10}).

The idea of the proof is to differentiate the Ernst potential with 
respect to $\xi$ and to use (\ref{heat1}) and (\ref{c10}) to relate 
these derivatives to directional derivatives of the theta functions. 
We get 
\ben
\left(\log\Ec\right)_\xi=
\frac{1}{4}\left\{D_\xi D_\xi\log\Ec+(D_\xi\log\Thho)^{2}\right.
\een
\be
\left.-(D_\xi\log\Thhd)^{2}-\co D_\xi\log\left(\Thho\Thhd\right)\right\}
\la{good}
\ee
The resulting expression can be simplified with the help of Fay's 
identities. It follows from Fay's identity (\ref{Fay1}) with $\z=\int_{\xi}^{\infty^-}$, $a=\infty^-$, $b=\xi$, $c=\infty^+$ that 
\footnote{It is worth noticing at this point that the action of the operator $D_\xi$ on the Ernst potential 
has a priori nothing to do with the partial derivative of the Ernst potential with respect to $\xi$: according to the
definition (\ref{defD}),  $D_\xi\Ec$ is just a directional derivative  of $\Ec$ with respect to
$\qb$ in the direction given by the values of  the basic holomorphic differentials at the branch point $\xi$ of the Riemann surface $\L$.}
\be
D_\xi\log\Ec = c_1 (\infty^-,\xi,\infty^+) + c_2(\infty^-,\xi,\infty^+)\f{\Thh^2(0)}{\Thhd\Thho}\;;
\la{good1}
\ee
applying the operator $D_\xi$ once more to both sides of this identity, we get
\ben
D_\xi D_\xi\log\Ec=\cd D_\xi\left\{\frac{\Thh^2(0)}{\Thhd\Thho}\right\}\;.
\een
Substituting this expression into (\ref{good}), we arrive at the formula
\ben
\left(\log\Ec\right)_\xi=\frac{1}{4}\cd D_\xi\left\{\frac{\Thh^2(0)}{\Thhd\Thho}\right\}
\een
\ben
+\frac{1}{4}D_\xi\log\left\{
\Thho\Thhd\right\}
\left\{D_\xi\log\Ec-\co\right\}\;.
\een

We use (\ref{good1}) again to 
simplify the last term. The result is
\be
\left(\log\Ec\right)_\xi=\frac{\cd}{2}\frac{\Thh(0) 
D_\xi\Thh(0)}{\Thhd\Thho},
\la{Exi}\ee
which is equivalent to  (\ref{Epoxi}).
The expression  (\ref{Epoxibar}) for $\Ec_\xibar$  can be proved analogously.
$\Box$

\subsection{Action of the Laplace operator on  the Ernst potential}
The same techniques can be used to determine the second derivatives 
of the Ernst potential which enter the axisymmetric Laplace operator.
\begin{theorem}
The action of the cylindrical  Laplace operator (\ref{delta0}) on the Ernst potential has the following form:
\be
\Delta\Ec =-2\cd c_2(\xi,\xibar,\infty^+) \frac{\Thhdb}{\Thh^3(\int_\xi^{\infty^-})} D_\xibar\Thhxi D_\xi\Thh(0)\;.
\la{Laplace}\ee
where the ratio of the prime-forms $c_2$ is defined  by  (\ref{c2}).
\end{theorem}

To prove  (\ref{Laplace}) we need to compute  the  derivatives with 
respect to $\xibar$ of all three 
multipliers in (\ref{Epoxi}) with the help of
the degenerated versions (\ref{Fay1}) and (\ref{Fay2}) of Fay's identities. 
These derivatives are given by the following three propositions.
\begin{proposition}\la{t1} The following identity holds:
\ben
4\left\{\log\frac{\Thh(0)}{\Thh(\int_\xi^{\infty^-})}\right\}_\xibar=
-c_2^2(\xi,\xibar,\infty^-)+c_1^2(\xi,\xibar,\infty^-)
\een
\be
-2c_2(\xi,\xibar,\infty^-) D_\xibar\log\Thhxi\frac{\Thhxi\Thhdb}
{\Thhd\Thh(0)}.
\la{derxib}\ee
\end{proposition}
{\it Proof.}
Using identity  (\ref{heat1}) as before,  we can write down the l.h.s. of 
(\ref{derxib}) as 
\ben
D_\xibar D_\xibar\log\frac{\Thh(0)}{\Thhd}+
D_\xibar\log\left\{\Thh(0)\Thhd\right\}D_\xibar\log\frac{\Thh(0)}{\Thhd}
\een
\ben
- \co D_\xibar\log\Thhd.
\een
Using the once degenerated Fay identity  (\ref{Fay1}) twice, we transform this expression to
\be
-\cdxi\frac{D_\xibar\left(\Thhxi\Thhdb\right)}{\Thh(0)\Thhd}
\la{hren}\ee
\ben
+\left(-c_1(\infty^-,\xibar,\infty^+)-\coxi\right)D_\xibar\log\Thhd-
\coxi D_\xibar\log\Thh(0).
\een
Since it follows directly from the definition (\ref{c1}) of the function $c_1$ that 
\be
c_1(\infty^-,\xibar,\infty^+)=-2\coxi\;,
\la{c11}\ee
the last two terms in (\ref{hren}) can be combined, which leads to 
\be
-\cdxi \frac{\Thhxi\Thhdb}{\Thh(0)\Thhd}\left(2D_\xibar\log\Thhxi
+D_\xibar\log\frac{\Thhdb}{\Thh(\int_\xibar^\xi)}\right)
\la{hren1}\ee
\ben
+\coxi D_\xibar\log\frac{\Thhd}{\Thh(0)}\;.
\een
We can use Fay's identities to further simplify (\ref{hren1}). The 
idea is to eliminate all derivatives of theta functions except of 
those with argument $\int_{\xi}^{\bar{\xi}}$.
For the second term in the first line we apply
(\ref{Fay1}) with $\z=\int_\xibar^\xi, a=\xi, b=\xibar, c=\infty^-$, for 
the last term we use the same identity with
$a=\xi,\; b=\xibar,\;c=\infty^-$. Subsequent simplification of the obtained  expression leads to (\ref{derxib}).
$\Box$
\vskip0.3cm
The next proposition gives the $\xibar$-derivative of the second multiplier in (\ref{Epoxi}):
\begin{proposition}\la{t2}
The following relation holds:
\be
2\left(D_\xi\log\Thh(0)\right)_\xibar=\dd\frac{\Thh(\int_\xibar^\xi)}{\Thh^2(0)}D_\xibar\Thhxi.
\la{fac2}\ee
\end{proposition}
{\it Proof.}  
Using (\ref{heat1}) and (\ref{Rauch1}), we get
\ben
\left(D_\xi\log\Thh(0)\right)_\xibar=\frac{1}{4} D_\xi\left(
D_\xibar D_\xibar\log\Thh(0)+\left(D_{\xibar}\log\Thh(0)\right)^{2}\right)\een
\ben
+\frac{1}{2}
\frac{W(\xi,\xibar)}{d\tau_\xi d\tau_\xibar}D_\xibar\log\Thh(0).
\een
Applying the twice degenerated Fay identity (\ref{Fay2}) and its 
$D_{\bar{\xi}}$-derivative to the different terms of  this expression, and
taking into account that 
$\frac{W(\xi,\xibar)}{d\tau_\xi d\tau_\xibar}=-d_1(\xi,\xibar)$, we transform this expression to 
the r.h.s. of (\ref{fac2}).
$\Box$
\vskip0.3cm
The subsequent statement provides the expression for the $\xibar$-derivative of the third term in (\ref{Epoxi}): 
\begin{proposition}\la{t3}
The following relation holds:
\be
 \partial_\xibar\log\cd=-\frac{1}{2}\left(c_1^2(\xi,\xibar,\infty^-)
-c_2^2(\xi,\xibar,\infty^-)\right)-\frac{1}{2(\xibar-\xi)}
\la{C2xibar}\ee
\end{proposition}
{\it Proof.}
In the proof we 
shall need a corollary of formula (\ref{root}):
\begin{lemma}
The following relation holds:
\be
\f{\pm 
1}{\sqrt{\xi-\xibar}}=
\frac{c_2(\infty^-,\xi,\infty^+)}{2Q}=c_{2}(\bar{\xi},\xi,\infty^{+})
\la{rootcol}\ee
The correct sign in (\ref{rootcol}) depends on the choice of all branches 
of the square roots in (\ref{rootcol}) and is unessential
for our purposes.  
\end{lemma}
{\it Proof.}
To prove (\ref{rootcol}) it is sufficient to consider the ratio of two 
root functions (\ref{root}): one  with
$\l_n=\xi$, $\l_m=\xibar$, and $a=\infty^+$ and another   with $\l_n=\xi$, $\l_m=\xibar$ and $a\to\l_m$.
Then the unknown function $C$ in (\ref{root}) drops out and we end up 
with (\ref{rootcol}). 
$\Box$

Relation  (\ref{rootcol}) implies 
\ben
\left(\log\cd)\right)_\xibar=\left(\log \realpart\right)_\xibar
+\frac{1}{2(\xi-\xibar)}.
\een
Now we shall prove that for the function $\realpart(\xi,\xibar)$ given by (\ref{C00})  
\be\la{Qxibar}
2\left(\log \realpart\right)_\xibar=c_2^2(\xi,\xibar,\infty^-)-c_1^2(\xi,\xibar,\infty^-).
 %c_2^2(\infty^-,\xi,\infty^+)-c_1^2(\infty^-,\xi,\infty^+).
\ee
It is convenient to use the representation of $\realpart$ in terms of  
theta functions with zero characteristics (\ref{C00}):
\ben
\left(\log \realpart\right)_\xibar
=\left(\log\frac{\Th(\int_\xi^{\infty^-})}{\Th(0)}\right)_\xibar+
\left(\log\frac{\Th(\int_\xibar^{\infty^-})}
{\Th(\int_\xi^\xibar)}\right)_\xibar.
\een
Using the result of proposition \ref{t1} with $\pb=\qb=0$, we see that 
\ben
\left(\log\frac{\Th(\int_\xi^{\infty^-})}{\Th(0)}\right)_\xibar= 
\frac{1}{4}\left(c_2^2(\xi,\xibar,\infty^-)-c_1^2(\xi,\xibar,\infty^-)\right),
\een
since $D_\xibar\log\Thxi$ vanishes being a directional derivative at zero of an even function. 
In the same way one can prove that
\be
\left(\log\frac{\Th(\int_\xibar^{\infty^-})}
{\Th(\int_\xi^\xibar)}\right)_\xibar=\frac{1}{4}\left(c_2^2(\xi,\xibar,
\infty^-)-c_1^2(\xi,\xibar,\infty^-)\right).
\la{above}\ee
$\Box$
\vskip0.3cm
Propositions \ref{t1}, \ref{t2} and  \ref{t3}
lead to (\ref{Laplace}) if we take into account the  next lemma:
\begin{lemma}
The following identity holds:
\be
\frac{1}{\xi-\xibar}\frac{c_2(\infty^-,\xibar,\infty^+)}{\cd}+\dd=0.
\la{root2}\ee
\end{lemma} 
{\it Proof.}
We rewrite the left hand side in prime forms, using (\ref{rootcol}) 
for $(\xi-\xibar)$.
Then
\ben
\frac{1}{\xi-\xibar}\frac{c_2(\infty^-,\xibar,\infty^+)}{\cd}=
\frac{E(\infty^+,\xibar)}{E^2(\xibar,\xi)E(\infty^+,\xi)d\tau_\xi}
\frac{E(\infty^-,\xi)}{E(\infty^-,\xibar)d\tau_\xibar}
\een
\be
=-\frac{1}{E^2(\xi,\xibar)d\tau_\xi d\tau_\xibar},
\la{minus}\ee
here we used that $\int_\xi^{\infty^+}=-\int_\xi^{\infty^-}$ and 
$\int_\xibar^{\infty^+}=-\int_\xibar^{\infty^-}$ and took into account
that the prime form is proportional to a theta-function with odd characteristic.
The minus sign in  (\ref{minus}) appears due to lemma \ref{stralem}.

\subsection{The Ernst equation}

To verify that (\ref{ES}) is a solution of the Ernst equation, one has to compare the 
action (\ref{Laplace}) of the Laplace operator on the 
Ernst potential with the expression
\be
\frac{8\Ec_\xi\Ec_\xibar}{\Ec+\bar{\Ec}}=\f{c_2(\infty^-,\xi,\infty^+)c_2(\infty^-,\xibar,\infty^+)}{\realpart}
\f{\Thhdb}{\Thh^3(\int_{\xi}^{\infty^-})} D_\xi \Thh(0)D_{\xibar}\Thhxi
\la{rhs}\ee
computed from (\ref{Epoxi}), (\ref{Epoxibar}) and (\ref{E+Ebar}).
The coincidence of these terms follows from the definitions of $c_2$ and $\realpart$.

\section{Metric functions for the stationary axisymmetric vacuum}

The metric of the stationary axisymmetric 
vacuum spacetimes can be written in the  Weyl--Lewis--Papapetrou form (see \cite{exac})
\begin{equation}\label{3.1}
\mathrm{ d} s^2 =-e^{2U}(\mathrm{ d} t+A\mathrm{ d} \phi)^2+e^{-2U}
\left(e^{2k}(\mathrm{ d} \rho^2+\mathrm{ d} \zeta^2)+
\rho^2\mathrm{ d} \phi^2\right)
\label{vac1}
\end{equation}
where $\rho$ and $\zeta$ are Weyl's canonical coordinates and 
$\partial_{t}$ and $\partial_{\phi}$ are the  commuting asymptotically
timelike respectively spacelike Killing vectors. 

In this case the vacuum field equations are equivalent 
to the Ernst equation (\ref{Ernstint}) for the 
complex potential $\Ec$. For a given  
Ernst potential, the  metric (\ref{vac1}) can be constructed as
follows: the metric function 
$e^{2U}$ is equal to the real part of the Ernst potential, which can be 
written in the form (\ref{E+Ebar}). The 
functions $A$ and $k$ can be obtained via a line integration from the equations
\begin{equation}
   A_{\xi}=2\rho\frac{(\Ec-\bar{\Ec})_{\xi}}{
    (\Ec+\bar{\Ec})^{2}}
    \label{axi},
\end{equation}
and 
\begin{equation}
    k_{\xi}=(\xi-\bar{\xi})
    \frac{\Ec_{\xi}\bar{\Ec}_{\xi}}{
    (\Ec+\bar{\Ec})^{2}}\;.
    \label{kxi}
\end{equation}

Explicit integration of equations (\ref{axi}) and (\ref{kxi}) is rather non-trivial; for the algebro-geometric solutions (\ref{solint}) it was carried out explicitly,  exploiting the
zero-curvature representation, in the papers 
\cite{Koro88,KorMat00,Klein96}.
In the sequel we show how to achieve these results on the sole base of Fay's identities and Rauch's formulas.

\subsection{Metric function $A$}

It was shown in 
\cite{Koro88} with the help of the inverse scattering method 
that the function $A$, corresponding to the Ernst potential (\ref{hyper}),  is related to a logarithmic 
derivative of theta functions which was  alternatively expressed  in 
\cite{Klein96} via 
theta functions themselves. One has the following
\begin{proposition}
    Let $A_{0}$ be a constant with respect to $\xi$ and $\bar{\xi}$.
    Then the metric function $Ae^{2U}$ for the  Ernst potential 
    (\ref{hyper}) is given by the expression:
\begin{equation}
    (A-A_{0})e^{2U}=-\rho\left(\f{1}{\realpart}\frac{\Thh(0)
    \Thh(\int_{\xi}^{\infty^{-}}+\int_{\bar{\xi}}^{\infty^{-}})}{
    \Thh(\int_{\xi}^{\infty^{-}})
    \Thh(\int_{\bar{\xi}}^{\infty^{-}})}-1\right)
    \label{a}.
\end{equation}
\end{proposition}
{\it Proof.}
    We have to show that equation (\ref{axi}) is satisfied with 
    the function $A$ given by 
    expression (\ref{a}). It is convenient to introduce the auxiliary 
    function  $Z:=(A-A_{0})e^{2U}$; then 
equation (\ref{axi}) is obviously equivalent to the equation  
    \begin{equation}
    Z_{\bar{\xi}}=\frac{1}{\mathcal{E}+\bar{\mathcal{E}}}\left((Z+\rho) 
    \bar{\Ec}_{\bar{\xi}}+ (Z-\rho)\mathcal{E}_{\bar{\xi}}\right)
    \label{Z1}.
\end{equation}
The first step in the proof is to establish the relation
\begin{eqnarray}
    4\left(\log 
    \frac{\Thh(\int_{\xi}^{\infty^{-}}+\int_{\bar{\xi}}^{\infty^{-}})}{
    \Theta(\int_{\bar{\xi}}^{\infty^{-}})}\right)_{\bar{\xi}}&=&
    -3c_{1}^{2}(\xi,\bar{\xi},\infty^{-})+
    c_{2}^{2}(\xi,\bar{\xi},\infty^{-})\left(
    4Q \frac{\Thh(\int_{\xi}^{\infty^{-}})\Thh(\int_{\bar{\xi}}^{\infty^{-}})}{
    \Thh(0)\Thh(\int_{\xi}^{\infty^{-}}+\int_{\bar{\xi}}^{\infty^{-}})}
    -1\right)
    \nonumber\\
    &&+2c_{2}(\xi,\bar{\xi},
    \infty^{-}) \frac{\Thh(\int_{\xi}^{\infty^{-}}) \Thh(\int_{\xi}^{\infty^{-}}
    +\int_{\xi}^{\infty^{-}})}{\Thh(\int_{\bar{\xi}}^{\infty^{-}})
    \Thh(\int_{\xi}^{\infty^{-}}+\int_{\bar{\xi}}^{\infty^{-}})}
    D_{\bar{\xi}}\log \Thh(0)
    \label{Zh1}.
\end{eqnarray}
The proof of this statement follows step by the step the proof of 
proposition \ref{t1}. Using (\ref{heat1}), we get for the l.h.s. of (\ref{Zh1})
\begin{eqnarray}
    D_{\bar{\xi}}D_{\bar{\xi}}\log \frac{\Thh(\int_{\xi}^{\infty^{-}}+\int_{\bar{\xi}}^{\infty^{-}})}{
    \Theta(\int_{\bar{\xi}}^{\infty^{-}})}+D_{\bar{\xi}}\log\frac{\Thh(\int_{\xi}^{\infty^{-}}+\int_{\bar{\xi}}^{\infty^{-}})}{
    \Theta(\int_{\bar{\xi}}^{\infty^{-}})}D_{\bar{\xi}}\log\left(
    \Thh(\int_{\xi}^{\infty^{-}}+\int_{\bar{\xi}}^{\infty^{-}})
    \Theta(\int_{\bar{\xi}}^{\infty^{-}})\right)\nonumber\\
    +c_{1}(\infty^{-},\bar{\xi},\infty^{+})\left( 2D_{\bar{\xi}}\log 
    \Thh(\int_{\xi}^{\infty^{-}}+\int_{\bar{\xi}}^{\infty^{-}})-
    D_{\bar{\xi}}\log\Theta(\int_{\bar{\xi}}^{\infty^{-}})\right)
    \label{Zh2}.
\end{eqnarray}
With the help of degenerated Fay's identity (\ref{Fay1}) with $a=\xi$, $b=\bar{\xi}$, 
$c=\infty^{-}$ and $\mathbf{z}=\int_{\bar{\xi}}^{\infty^{-}}$, its 
$D_{\bar{\xi}}$ derivative and the formula (\ref{c11}), we can rewrite the expression (\ref{Zh2}) as follows:
\begin{eqnarray}
     &  & c_{2}(\xi,\bar{\xi},\infty^{-})\frac{\Thh(\int_{\xi}^{\infty^{-}})
     \Thh(\int_{\xi}^{\infty^{-}}+\int_{\xi}^{\infty^{-}})}{
     \Thh(\int_{\bar{\xi}}^{\infty^{-}})\Thh(\int_{\xi}^{\infty^{-}}
     +\int_{\bar{\xi}}^{\infty^{-}})}D_{\bar{\xi}}\log\left(\Thh(\int_{\xi}^{\infty^{-}})
     \Thh(2\int_{\xi}^{\infty^{-}})\right)
    \nonumber  \\
     &  & -3c_{1}(\xi,\bar{\xi},\infty^{-}) \left(
     c_{1}(\xi,\bar{\xi},\infty^{-})+c_{2}(\xi,\bar{\xi},\infty^{-})
     \frac{\Thh(\int_{\xi}^{\infty^{-}})\Thh(\int_{\xi}^{\infty^{-}}+
     \int_{\xi}^{\infty^{-}})}{\Thh(\int_{\bar{\xi}}^{\infty^{-}})
     \Thh(\int_{\xi}^{\infty^{-}}+\int_{\bar{\xi}}^{\infty^{-}})}\right)
    \label{Zh3}.
\end{eqnarray}
The theta derivatives in the first line of (\ref{Zh3}) can be related 
to derivatives of the theta function with zero argument via degenerated Fay's identity
(\ref{Fay1}) for $a=\xi$, $b=\bar{\xi}$, $c=\infty^{-}$, 
$\mathbf{z}=0$ and $a=\infty^{+}$, $b=\bar{\xi}$, $c=\infty^{-}$, 
$\mathbf{z}=0$ respectively (note that $\int_{\xi}^{\infty^{-}}+
     \int_{\xi}^{\infty^{-}}=\int_{\infty^{+}}^{\infty^{-}}$). 
     The resulting expression can be 
simplified using (\ref{rootcol}) and Fay's identity (\ref{Fay1}) with 
$\zb=\int_\xi^{\infty^-}$, $a=\xi$, $b=\xibar$, $c=\infty^-$, 
$d=\infty^+$,
\be
    \Thh(\int_{\xi}^{\infty^{-}}+\int_{\bar{\xi}}^{\infty^{-}})
  \Thh(0)+\Thh(2\int_{\xi}^{\infty^{-}})\Thh
    (\int_{\xi}^{\bar{\xi}})
= 2\realpart\Thh(\int_{\xi}^{\infty^{-}})
\Thh(\int_{\bar{\xi}}^{\infty^{-}})\;,
    \label{Z4}
\ee
to (\ref{Zh1}). 

Furthermore, with the help of relations   (\ref{derxib}) 
and (\ref{Qxibar}), we transform this expression as follows:
\begin{eqnarray}
    \log \left(\frac{\Thh(0)
    \Thh(\int_{\xi}^{\infty^{-}}+\int_{\bar{\xi}}^{\infty^{-}})}{
    \realpart\Thh(\int_{\xi}^{\infty^{-}})
    \Thh(\int_{\bar{\xi}}^{\infty^{-}})}\right)_{\bar{\xi}}= 
    c_{2}^{2}(\xi,\bar{\xi},\infty^{-})\left(
    Q 
    \frac{\Thh(\int_{\xi}^{\infty^{-}})\Thh(\int_{\bar{\xi}}^{\infty^{-}})}{
    \Thh(0)\Thh(\int_{\xi}^{\infty^{-}}+\int_{\bar{\xi}}^{\infty^{-}})}
    -1\right)
    \nonumber\\
    +\frac{c_{2}(\xi,\bar{\xi},
        \infty^{-})}{2} \frac{\Thh(\int_{\xi}^{\infty^{-}}) \Thh(\int_{\xi}^{\infty^{-}}
    +\int_{\xi}^{\infty^{-}})}{\Thh(\int_{\bar{\xi}}^{\infty^{-}})
    \Thh(\int_{\xi}^{\infty^{-}}+\int_{\bar{\xi}}^{\infty^{-}})}
    D_{\bar{\xi}}\log \Thh(0)\nonumber\\
    -\frac{c_{2}(\xi,\bar{\xi},
        \infty^{-})}{2} \frac{\Thh(\int_{\xi}^{\bar{\xi}}) 
	\Thh(\int_{\bar{\xi}}^{\infty^{-}})}{\Thh(0)
    \Thh(\int_{\xi}^{\infty^{-}})}D_{\bar{\xi}}\log \Thh(\int_{\xi}^{\bar{\xi}})
    \label{Zh5}.
\end{eqnarray}
Taking into account the  relation (\ref{rootcol}), this implies the following relation:
\begin{eqnarray}
    Z_{\bar{\xi}}&=&\frac{c_{2}(\xi,\bar{\xi},
        \infty^{-})\rho}{2Q} \frac{\Thh(0) \Thh(\int_{\xi}^{\infty^{-}}
    +\int_{\xi}^{\infty^{-}})}{\Thh^{2}(\int_{\bar{\xi}}^{\infty^{-}})
    }
    D_{\bar{\xi}}\log \Thh(0)\nonumber\\
    &&-\frac{c_{2}(\xi,\bar{\xi},
        \infty^{-})\rho}{2Q} \frac{\Thh(\int_{\xi}^{\bar{\xi}}) 
	\Thh(\int_{\bar{\xi}}^{\infty^{-}}+\int_{\xi}^{\infty^{-}})}{
    \Thh^{2}(\int_{\xi}^{\infty^{-}})}D_{\bar{\xi}}\log \Thh(\int_{\xi}^{\bar{\xi}})
    \label{Zh6}.
\end{eqnarray}
Whereas the expression for $Z-\rho$ follows directly from (\ref{a}), we can write 
$Z+\rho$, using Fay's identity (\ref{Z4}), in the convenient form
\begin{equation}
    Z+\rho = \f{\rho}{Q}
    \frac{\Thh(2\int_{\xi}^{\infty^{-}})\Thh
    (\int_{\xi}^{\bar{\xi}})}{\Thh(\int_{\xi}^{\infty^{-}})
\Thh(\int_{\bar{\xi}}^{\infty^{-}})}
    \label{Z6}.
\end{equation}
 
Relation (\ref{Zh6}) turns out to be equivalent to (\ref{Z1}) if we use equalities 
(\ref{rootcol}), (\ref{E+Ebar}), (\ref{Epoxi}) and (\ref{Epoxibar}).
$\Box$

\subsection{Metric function $e^{2k}$}

The metric function $e^{2k}$ was calculated in \cite{KorMat00} as the 
$\tau$-function of the Schlesinger system associated to the Ernst 
equation. Here we shall prove the resulting formula
using Fay's identities. 
\begin{theorem}
The metric function $e^{2k}$ is given by 
\begin{equation}
    e^{2k}= K \frac{\Thh(0)
    \Thh(\int_{\xi}^{\bar{\xi}})}{\Theta(0)
    \Theta(\int_{\xi}^{\bar{\xi}})}.
    \label{k}
\end{equation}
where $K$ is a constant, and where as before $ \int_{\xi}^{\bar{\xi}}\equiv -\f{1}{2}(1,\dots,1)$.
 \end{theorem}
{\it Proof.}
We have to show that (\ref{kxi}) is satisfied with $k$ given by
(\ref{k}).  Taking into account the relations (\ref{Epoxi}), (\ref{Epoxibar}), 
and (\ref{rootcol}),
we obtain the following proposition we need to prove:

\begin{proposition}
The following identity holds:
\be
\frac{1}{8}\left(D_{\xi}D_{\xi}\log\frac{\Thh(0)\Thhxi}{\Th(0)\Thxi}+(D_\xi\log\Thh(0))^2+(D_\xi\log\Thhxi)^2\right)
\label{F1}\ee
\ben
=\frac{1}{4}D_\xi\log\Thh(0)D_\xi\log\Thhxi.
\een
\end{proposition}
{\it Proof.}
As  the first step of the proof of identity (\ref{F1}) we observe that (\ref{F1}) can be rewritten
in terms of the theta-function without characteristics as follows:
\be
D_{\xi}D_{\xi}\log\frac{\Th(\V)\Th(\int_\xi^\xibar+\V)}{\Th(0)\Thxi
}+(D_\xi\log\Th(\V))^2+(D_\xi\log\Th(\int_\xi^\xibar+\V))^2
\la{F2}\ee
\ben
=2D_\xi\log\Th(\V)D_\xi\log\Th(\int_\xi^\xibar+\V),
\een
where $\V\equiv\B\pb+\qb$
i.e. all exponential terms arising from relation (\ref{thchar}) between the theta-function with characteristics and the theta-function
with shifted argument drop out; therefore the statement  (\ref{F1}) takes the form (\ref{F2}).

The idea of the proof of identity (\ref{F2})  is the 
following: we define a  function $F$ as the difference of the left-hand 
and the right-hand side of (\ref{F2}).
We show that the  derivatives of the function $F$ with respect to any components
 $p_\alpha$ and any $q_\alpha$ of vectors $\pb$ and $\qb$ vanish.
Then function $F$ must be a constant with respect to $\pb$ and $\qb$;  thus it is sufficient to observe
that this function vanishes  at $\pb=\qb=0$. 

Function $F$ depends only on the  combination $\V\equiv\B\pb+\qb$; therefore, all partial derivatives of $F$ with respect to each $p_\a$
are linear combinations of the partial derivatives with respect to $q_\a$; thus it is sufficient to prove that 
all partial derivatives of $F$ with respect to $q_\alpha$ vanish.

In turn, to show that all partial derivatives of $F$ with respect to $q_\alpha$ are equal to zero, it is sufficient to prove
that  $D_{b}F\equiv \sum_{\a=1}^g\f{\p F}{\p q_\a} \o_\a(b)/d\tau_{b}
$ vanishes for an arbitrary point $b\in\mathcal{L} $, 
taking into account the following lemma:

\begin{lemma}
There exists a positive divisor $b_1+\dots + b_g$  of degree $g$ on $\L$ such that
the vectors\\ $\o(b_1)/d\tau_{b_1}\;,\dots,\o(b_g)/d\tau_{b_g}$
are linearly independent.
\end{lemma}
{\it Proof.}
Suppose the opposite, i.e.\ that $\det\{\o_\a(b_\beta)\}$ vanishes for any divisor $b_1+\dots + b_g$.
Let us integrate this determinant along a basic cycle $a_\beta$ with respect to variable $b_\beta$ for each $\beta$.
On one hand, the result should equal $0$ according to our assumption.
On the other hand, we get the determinant of the unit matrix, which
equals $1$. This contradiction proves the lemma.
\footnote{It was noticed by the  referee that this lemma also  has a  geometrical
interpretation: it means  that the canonical model of the curve is not contained in any hyperplane.}
$\Box$

Thus for suitably chosen $b$, the vector $\omega(b)/d\tau_{b}$ will 
take all values in $\C^{g}$. If one can show that $D_{b}F=0$ for 
arbitrary $b$, this implies that $F$ must be a constant.  

Now  let us  calculate the $D_{b}$ derivative of $F$  (\ref{F2}) where 
$b\neq\xi$ is an otherwise arbitrary point on $\mathcal{L}$. With the 
help of Fay's identity  (\ref{Fay2}) we can write down this derivative as follows:
\begin{eqnarray}
    D_{b}F&=& d_{2}(b,\xi)\frac{\Th(\int_{\xi}^{b}+\V)\Th(\int_{b}^{\xi}+\V)}{
    \Th^{2}(\V)}D_{\xi}\ln \frac{\Th(\int_{\xi}^{b}+\V)
        \Th(\int_{b}^{\xi}+\V)}{\Th^{2}(\int_{\xi}^{\bar{\xi}}+\V)}
    \nonumber  \\
    &&+d_{2}(b,\xi)\frac{\Th(\int_{\bar{\xi}}^{b}+\V)
    \Th(\int_{b}^{\bar{\xi}}+\V)}{
    \Th^{2}(\int_{\xi}^{\bar{\xi}}+\V)}
    D_{\xi}\ln\frac{\Th(\int_{\bar{\xi}}^{b}+\V)
        \Th(\int_{b}^{\bar{\xi}}+\V)}{\Th^{2}(\V)}
    \label{k2}.
\end{eqnarray}
The degenerated Fay identity (\ref{Fay1}) implies
\be
    D_{\xi}\ln \left\{\Th(\int_{\xi}^{b}+\V)
    \Th(\int_{b}^{\xi}+\V)\right\}
\la{k2a}\ee
\ben
=2D_{\xi}\ln\Th(\int_{\xi}^{\bar{\xi}}+\V)+
    c_{2}(\bar{\xi},\xi,b)\frac{\Th(\V)}{\Th(\int_{\xi}^{\bar{\xi}}+\V)
    }\left(\frac{\Th(\int_{\bar{\xi}}^{b}+\V)}{\Th(\int_{\xi}^{b}+\V)}
    -\frac{\Th(\int_{b}^{\bar{\xi}}+\V)}{\Th(\int_{b}^{\xi}+\V)}\right).
\een
Substituting (\ref{k2a}), together with  the 
corresponding relation for $D_{\xi}\{\Th(\int_{\bar{\xi}}^{b}+\V)
        \Th(\int_{b}^{\bar{\xi}}+\V)\}$, into (\ref{k2}),  we find that the $D_{b}$ 
derivative of $F$ is identically zero for 
all $b\neq \xi$.  Consequently, the difference  $F$ between the r.h.s. and l.h.s of (\ref{F2}) must be a constant 
with respect to the characteristics  $[\pb,\qb]$.
Considering the case $[\pb,\qb]=[0,0]$ we see that 
both sides of (\ref{kxi}) are zero in this case. This completes the proof.
$\Box$
$\Box$

\section{Outlook}\label{sec.5}
In this paper we have presented a unified approach to theta 
functional solutions to the Ernst equation, i.e.\  to the 
stationary axisymmetric vacuum Einstein equations. Based on Fay's trisecant 
identity, its degenerations and Rauch's variational formulas for
hyperelliptic Riemann surfaces, we proved the validity for formulas 
for the Ernst potential. The complete metric and the Ernst potential 
can be given explicitly in terms of theta functions. This explicit 
form free of derivatives of the metric made it 
possible in \cite{Klein00} to solve a boundary value problem for a 
relativistic dust disc in terms of a theta-functional Ernst 
potential. The description of the dust discs requires partial degeneration of the curve $\L$
and subsequent "condensation" of the double points, as was done in \cite{KorMat00}.

It is an open questions whether the methods outlined in this article 
can also be of direct use in the  solution of boundary value problems 
as e.g.\ in the context of dust discs or black-hole disc systems. 
It would be interesting
to extend this approach to the Einstein-Maxwell case where the 
theta-functional solutions are given on non-hyperelliptic Riemann 
surfaces (see \cite{Koro88}).

\textbf{Acknowledgement}\\
CK thanks for financial support by the Schloessmann foundation. Research of DK was supported by the grant of
Fonds pour la Formation de Chercheurs et l'Aide a la Recherche de 
Quebec and the grant of Natural Sciences and Engineering Research Council 
of Canada. We thank the anonimous referee for several useful suggestions.

\end{document}